\documentclass[reprint,pra,twocolumn,a4paper,superscriptaddress,aps]{revtex4}

\usepackage{graphicx}
\usepackage{dcolumn}
\usepackage{amsmath,amssymb,graphicx,color,dsfont}
\usepackage{mathtools,amssymb}
\usepackage{pdfpages}
\usepackage{bm}
\allowdisplaybreaks
\usepackage{tabularx}
\pdfoutput=1
\usepackage{relsize}
\usepackage[toc]{appendix}
\usepackage{upgreek}
\usepackage[utf8]{inputenc} 

\newcommand{\D}{$^{\circ}$}

\begin{document}

\title{
Tolerance analysis of non-depolarizing double-pass polarimetry
}

\author{Yimin Yu}
\email[]{yuym3133@smee.com.cn}
\thanks{all authors contributed equally.}
\affiliation{Shanghai Micro Electronics Equipment (Group) Co., Ltd, 1525 Zhangdong Road, Shanghai 201203, China}

\author{Nabila Baba-Ali}
\affiliation{Stamford, Connecticut, USA}

\author{Gregg M Gallatin}
\affiliation{Stamford, Connecticut, USA}
\date{\today}

\begin{abstract}
Double-pass polarimetry measures the polarization properties of a sample over a range of polar angles and all azimuths. Here, we present a tolerance analysis of all the optical elements in both the calibration and measurement procedures to predict the sensitivities of the {double-pass} polarimeter. 
The calibration procedure is described by a Mueller matrix based on the eigenvalue calibration method (ECM) \cite{Compain99}. Our numerical results from the calibration and measurement in the Mueller matrix description with tolerances limited by systematic and stochastic noise from specifications of commercially available hardware components are in good agreement with previous experimental observations. Furthermore, by using the orientation Zernike polynomials (OZP) which are an extension of the Jones matrix formalism, similar to the Zernike polynomials wavefront expansion, the pupil distribution of the polarization properties of non-depolarizing samples under test are expanded. Using polar angles ranging up to 25\D, we predict a sensitivity of 0.5\% for diattenuation and 0.3\D\ for retardance using the root mean square (RMS) of the corresponding OZP coefficients as a measure of the error. This numerical tool provides an approach for further improving the sensitivities of polarimeters via error budgeting and replacing sensitive components with those having better precision.
\end{abstract}

\maketitle



\section{Introduction}
Polarimeters characterize the polarization properties of materials.  They find application in, for instance, optical samples \cite{Ibrahim2009}, cancer non-invasive screening tools \cite{advanced2013} in clinics, hyper-numerical-aperture lithography \cite{PSM2005,Geh2007,Nomura2007,Nomura2008,Nomura2010} where controlled polarization enhances the contrast and thus enabling smaller structures to be written on the wafer.

Inherited from standard interferometry \cite{Goodwin2006}, the double-pass configuration detecting the phase shift between its two arms has been developed for sensing applications such as dilatometric measurement \cite{Ren2008} and pH monitoring \cite{Tou2014}. In polarimetry, a double-pass layout enables angle-resolved measurements, whereby the polarization response of a sample for a range of polar angles and all azimuths can be measured in a synchronous approach. This simplifies the measurement setup and saves time compared to, otherwise, an apparatus with a function of rotating a solid angle over a certain range. Since the light is transmitted through the sample being tested twice, each ray nominally picks up the same polarization properties in both the outgoing and return paths. Given the same apparatus errors outside the sample being tested, the double-pass configuration offers double the sensitivity of the polarization properties. The interferometric merit of the double-pass, on the other hand, is utilized in aligning the optical components in the angle-resolved polarimetry. While experimental demonstrations have validated the concept of double-pass polarimetry in angle-resolved polarization measurements \cite{Ibrahim2009}, repeatability analysis to tolerances of the double-pass polarimeter has not been studied systematically. The present work attempts to fill this gap by providing a detailed sensitivity analysis of the polarimeter repeatability.

\begin{figure*}[!hptb]
\begin{center}
\includegraphics[width=1.6 \columnwidth]{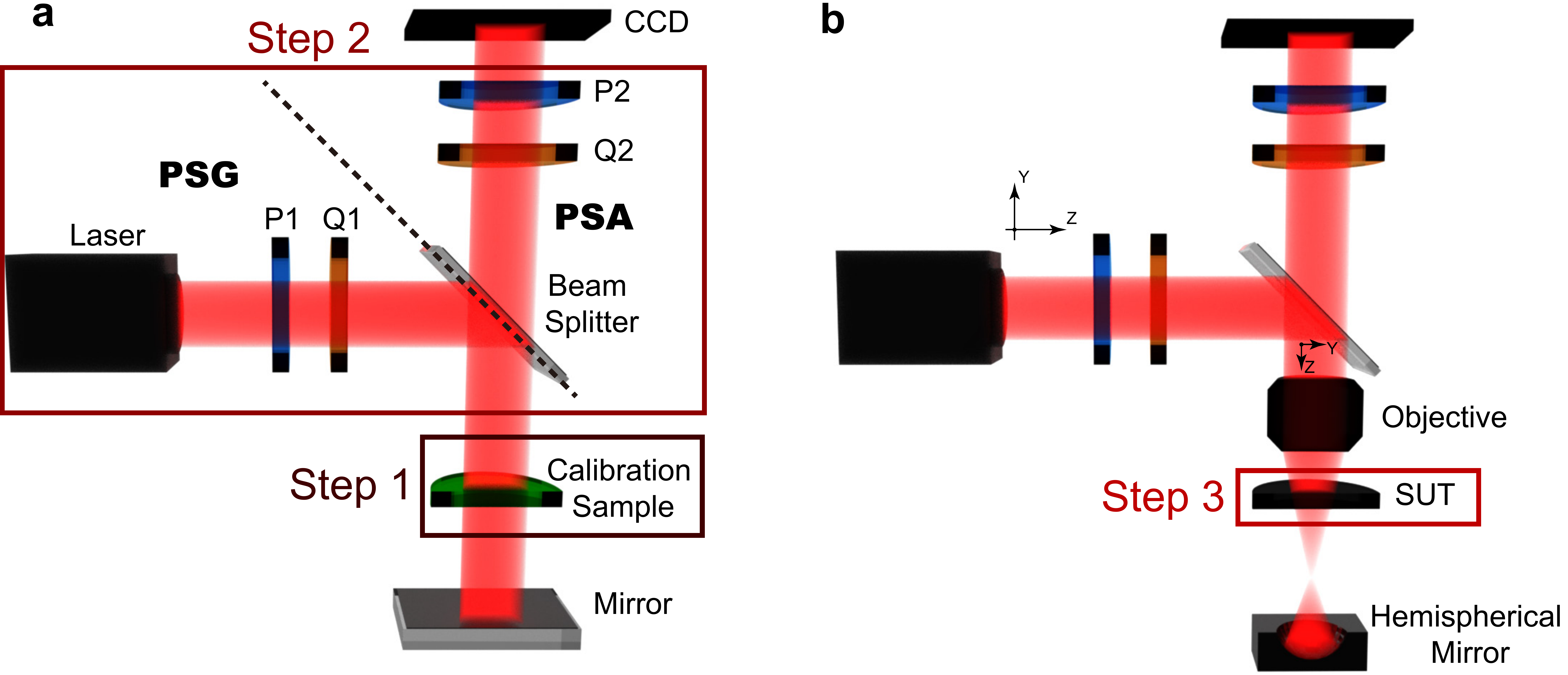}
\end{center}
  \caption{ Sketch of the polarimeter apparatuses for the calibration and measurement procedures. {\bf a} The setup for calibrating the polarization state generator (PSG) and polarization state analyzer (PSA). The PSG includes the coherent laser, a polarizer (P1), a quarter-wave plate (Q1), and the reflective path of the non-polarizing beam splitter (BS) successively along the optical path. The PSA has components of the transmission path of the BS, a quarter-wave plate (Q2) and a polarizer (P2). In step 1, the polarization properties of the calibration samples are extracted, serving to calibrate matrices for the PSG and PSA in step 2. {\bf b} In the apparatus for measuring a sample under test (SUT), the calibration samples and the mirror are replaced with an objective, the SUT and a hemispherical mirror to retro back the impinging beam.}
\label{Fig:layout}
\end{figure*}

An example of the operation of a double-pass polarimeter includes the calibration and measurement procedures. In the calibration apparatus as illustrated in Fig.~\ref{Fig:layout}a, a coherent laser illuminates a polarizer (P1) and a quarter-wave plate (Q1) successively before being reflected by a non-polarizing beam splitter (BS). The coherent laser, the polarization components P1 and Q1, together with the reflective path of the BS, form the polarization state generator (PSG). The laser beam then passes through the calibration sample in the forward and reverse directions with the help of a mirror. The change of the polarization state of the beam caused by the calibration sample and the mirror is analyzed by the polarization state analyzer (PSA) and readout from the CCD. The PSA consists of the transmission path of the BS, the quarter-wave plate Q2 and the polarizer P2. The goal of the calibration setup is to characterize the polarization properties of the PSG and PSA accurately using calibration samples and the eigenvalue calibration method (ECM) \cite{Compain99}. The polarization properties of the calibration samples can be extracted using the same setup. In the measurement procedure the calibration samples and the mirror are subsequently replaced with an objective lens, the sample under test (SUT) and a hemispherical mirror as shown in Fig.~\ref{Fig:layout}b. The focus of the laser beam from the objective is aligned to coincide with the center of the curvature of the hemispherical mirror, to ensure that the beam is reflected back along the incoming optical path. The SUT is placed away from the focus for the laser beam to cover its pupil.

In this work, we break down the angle-resolved measurement of a SUT into 3 steps. In step 1, the transmittance amplitude for the two orthogonal polarization eigenstates and the retardance of the calibration samples are extracted from the calibration apparatus by comparing the intensities with and without the calibration samples. Step 2 is an algorithmic procedure used to obtain the polarization properties of the PSG and PSA. This algorithm depends not only on the measured intensities with and without the calibration samples, but also on the polarization properties extracted in step 1. In step 3, the measurement setup employs the calibrated PSG and PSA to measure the polarization of the objective, SUT and hemispherical mirror together. Because the PSG and PSA are maintained unchanged during calibration and measurement, they cause no systematic change in the errors in measuring the SUT. Tolerance analysis of the components affecting the polarization from steps 1 to 3 results in the polarization measurement sensitivities. 

We characterize the polarization of a non-depolarizing sample in terms of its diattenuation and retardance, which quantify the transmission amplitude difference between the two orthogonal brightest and darkest axes and the phase difference between the two orthogonal fastest and slowest axes, respectively. For non-depolarizing samples, the Jones matrix representation of the polarization is all that is required and is simpler than the Mueller matrix representation, in that the Jones matrix uses fewer parameters, only 4 complex elements compared to 16 real elements for the Mueller matrix. The diattenuation and retardance across the pupil can be expanded in terms of the orientation Zernike polynomials (OZP) based on the Jones matrix formalism \cite{Tilmann2008,Ruoff2009,Ruoff2010}, and the RMS of the coefficients quantifies the diattenuation or retardance across the entire pupil by analogy with Zernike polynomials for wavefront expansion. By inputting tolerances of available commercial products into the numerical model, we predict a sensitivity of 0.5\%  RMS OZP  (a unit stands for the RMS of the corresponding OZP coefficients) for a diattenuation pupil, equivalent to a pupil with a mean diattenuation of 1\%. Likewise, the prediction of the sensitivity for a retardance pupil is 0.3\D\ RMS OZP corresponding to a pupil with a mean retardance of 0.6\D. 

This numerical tool takes the systematic and stochastic errors of each component in the system for both the calibration and measurement as inputs, and derives the sensitivities of diattenuation and retardance to errors in the measured values in a bottom-up approach. Whereas double-pass polarimeters can find application for characterizing incident-angle dependent variable attenuators \cite{VA2008}, wide-view-angle polarizers and retarders \cite{Nomura2007,Nomura2010,Yeh1982} in lithographic equipment, this numerical tolerance analysis paves the way for predicting the sensitivity of the polarization properties for those optical components. Furthermore, this numerical tool can help to improve sensitivity via error-budgeting \cite{incose2006}. Depending on the relative contribution of each tolerance error, targeted hardware could be replaced to improve the sensitivity.


\section{Methods}
\subsection{Step 1: Determining the properties of calibration samples}
Classical calibration procedures usually rely on standard samples with well-known properties \cite{Jell1997} or similar devices with higher accuracy. The former approach requires strict sample fabrication, while the later one limits the accuracy of the polarimeters to be calibrated to, roughly, that of the calibrating polarimeter. The ECM developed by Compain  et  al. \cite{Compain99} largely  relaxes  the requirement for special  calibration  samples,  and is able to extract the polarization properties of the calibration sample from the polarimeter itself, hence nominally guaranteeing measurement accuracy. The ECM uses linear dichroic polarizers and retarders with retardation far from 180\D\cite{Compain99,Stabo2009}. These polarization elements need to be homogeneous \cite{Lu1994}. That is their eigen polarization states of polarizing elements are orthogonal. Here we extend the ECM to double-pass polarimetry. Due to the flat mirror in the double-pass layout sketched in Fig.~\ref{Fig:layout}a, wave plates with retardance of 90\D are excluded from use as calibration samples. A dichroic polarizer and a 1/6-wave plate are selected as calibration samples in this work. 

Intensities modulated by the PSG and PSA are recorded. The calibration sample is first retracted from the optical path in the setup in Fig.~\ref{Fig:layout}a, leaving only the mirror. This results in the intensity matrix $i_0$
\begin{equation}
\hspace{-5mm}
\begin{split}
\underbrace{\begin{bmatrix}
i_0^{1,1} & i_0^{1,2} & \hdots & i_0^{1,v} \\
i_0^{2,1} & i_0^{2,2} & \hdots & i_0^{2,v}  \\
\vdots & \vdots &\ddots & \vdots\\
i_0^{u,1} & i_0^{u,2} & \hdots & i_0^{u,v}
\end{bmatrix}}_{i_0}
=
\underbrace{\begin{bmatrix}
a^1_{1,1} & a^1_{1,2} &  a^1_{1,3} & a^1_{1,4} \\
a^2_{1,1} & a^2_{1,2} &  a^2_{1,3} & a^2_{1,4}  \\
\vdots & \vdots &\vdots & \vdots\\
a^u_{1,1} & a^u_{1,2} &  a^u_{1,3} & a^u_{1,4}
\end{bmatrix}}_{a} \times \\
m_{\rm mirror} \times 
\underbrace{\begin{bmatrix}
w_0^1 & w_0^2 & \hdots & w_0^v \\
w_1^1 & w_1^2 & \hdots & w_1^v   \\
w_2^1 & w_2^2 & \hdots & w_2^v \\
w_3^1 & w_3^2 & \hdots & w_3^v 
\end{bmatrix}}_{w}.
\end{split}
\label{i0matrix}
\end{equation}
Here matrix $a$ is the calculated PSA matrix from the intensity measurement. It is constructed from the 1\textsuperscript{st} to the $u$\textsuperscript{th} configuration of the PSA, using the first row of the Mueller matrices of the PSA. The calculated PSG matrix $w$ is formed by $v$ different configurations of Stokes vectors. The middle term $m_{\rm mirror}$  on the right hand side (RHS) of Eq.~\eqref{i0matrix} is the measured Mueller matrix for the mirror.

We then insert the dichroic polarizer and the 1/6-wave plate separately to obtain the intensity matrices 
\begin{equation}
i_i=am^b_{i}m_{\rm mirror}m^f_{i}w,
\label{ii}
\end{equation}
in which the subscript stands for the $i$\textsuperscript{th} calibration sample. Matrices $m^f_i$ and $m^b_i$ can be further decomposed to $m^f_{i} = R(\theta)m_iR(-\theta')$ and  $m^b_{i} = R(-\theta')m_iR(\theta)$, where $R(\theta)$ is the rotation matrix corresponding to azimuthal rotation angle $\theta$ of the calibration samples. The superscript $f$ and $b$ denote that the light passes through the calibration sample in a forward path and a backward path after reflection from the mirror, respectively. The measured Mueller matrix of the dichroic polarization elements $m_i$, with zero azimuthal angle, can be expressed as \cite{Compain99}
\begin{equation}
m_i = 
\begin{bmatrix}
t_X^2+t_Y^2 & t_X^2-t_Y^2 & 0 & 0 \\
t_X^2-t_Y^2 & t_X^2+t_Y^2 & 0 & 0  \\
0           & 0           & 2t_Xt_Y\cos\phi & 2t_Xt_Y\sin\phi\\
0           & 0           & -2t_Xt_Y\sin\phi & 2t_Xt_Y\cos\phi
\end{bmatrix},
\label{mi}
\end{equation}
in which $t_X$ and $t_Y$ are the measured transmittance amplitudes of the sample along the two orthogonal directions, X and Y. 
{We define the Z direction of the coordinate to be aligned with the ray propagation direction, the X direction to be pointing inside, and the Y direction to be pointing upwards at the start of the beam near the laser as demonstrated in Fig.~\ref{Fig:layout}b.}
The measured retardance difference between the X and Y directions is $\phi$.

The quotient matrix $c_i$ is defined as the product of the inverse of the intensity matrix $i_{\rm 0}$ and the matrix $i_{i}$, which gives
\begin{align}
\begin{split}
c_i & = i^{-1}_0 i_i=w^{-1}m^{-1}_{\rm mirror}m^b_im_{\rm mirror}m^f_iw\\
& \approx [R(-\theta)w]^{-1}[m_i]^2[R(-\theta)w].
\end{split}
\label{ci}
\end{align}
{The Mueller matrix of the mirror $m_{\rm mirror}$ is in the form of Eq.~\eqref{mi}, where the non-identity of the reflectance is expressed by the transmittance amplitudes $t_X$ and $t_Y$, and the retardance $\phi$ of the mirror in Eq.~\eqref{mi} is taken to be the sum of 180\D\ and noises.}
The last relation ($\approx$) becomes an equality when no noise is present in the measured intensity matrices $i_0,\   i_1$ or in the control of the azimuthal angle $\theta$ of the calibration samples. To ensure the uniqueness of the solutions, the full rank of the PSG's $w$ matrix is required for the inversion in Eq.~\eqref{ci} and the PSA’s matrix $a$ has the same requirement. The true combinations of the azimuthal angles of the polarizing elements in the PSG are chosen to maximize the absolute value of the determinant of the true PSG matrix $W$ in order to minimize the inversion error of $W$ in calculating the Mueller matrix for the calibration sample. Here, we use the convention that matrices with uncapitalized and capitalized letters symbolize the measured (or calculated) values and the actual (or true) values, respectively. The true PSA matrix $A$ is optimized in the same way. The coherent laser source beam in the PSG is modeled as a linearly polarized electrical field of $E_{in} = [1; 1]/\sqrt{2}$. The PSG uses 4 configurations in our simulations for convenience in performing the inversions, i.e., set $v=4$ in Eq.~\eqref{i0matrix}. Each configuration is obtained by varying the azimuthal angles of the polarizer P1 and the quarter-wave plate Q1. The true values of the polarization properties of the PSG and PSA used in the simulation are summarized in Tab.~\ref{Tab:WAcoef}.
\begin{table}[h!]
\caption{The true values of the polarization properties of the PSA and PSG.}
      \begin{tabular}{ccccc}
        \hline
           & PSG P1  & PSG Q1   & PSG BS$_{\rm reflect}$\\ \hline
        $T_X / R_X$ & 0.9 & 0.98 & $\sqrt{0.5}$\\
        $T_Y / R_Y$ & 0.0098 & 0.97  & $\sqrt{0.5}$\\
        $\Theta$ & 21\D, 63\D, 45\D, 52\D\  & -90\D, -88\D, 48\D, -3\D\   & 0\D \\ 
        $\Phi$ & 0\D  & 90\D   & 180\D \\ \hline
      \end{tabular}
      
      \begin{tabular}{ccccc}
        \hline
           & PSA BS$_{\rm transmit}$   & PSA Q2 & PSA P2\\ \hline
        $T_X$ & $\sqrt{0.5}$ & 0.98 & 0.9\\
        $T_Y$ & $\sqrt{0.5}$  & 0.97 & 0.0098\\
        $\Theta$ & 0\D\    & 40\D, -84\D, -46\D, 75\D\  & -39\D, 81\D, -9\D, -1\D \\ 
        $\Phi$ & 0\D   & 90\D & 0\D \\ \hline
      \end{tabular}
      \label{Tab:WAcoef}
\end{table}

The maximum absolute value of the determinant of the PSG is optimized to $\left|{\rm det}W\right|$=0.58 and that of the PSA is  $\left|{\rm det}A\right|$=0.06. The reflectance and transmittance amplitudes of the BS are idealized to be $\sqrt{0.5}$ in this modeling. Note that the Mueller matrix of the BS only affects the optimization of the azimuthal angle configurations for P1, Q1, Q2 and P2. It has no influence on the calibration error for the PSG $\Delta W = w - W$ or that for the PSA $\Delta A =a-A$ in step 2. In the experiments, the Mueller matrix for the transmission and reflection paths through the BS could be measured in advance using a single pass polarimeter in transmission \cite{Stabo2009,Korger2013} and reflection \cite{Compain99} to ensure the calibration accuracy. 

The quotient matrix $c_i$ in Eq.~\eqref{ci} is similar, in the linear algebra sense, to the square of the measured Mueller matrix of the calibration sample $[m_i]^2$ given that matrix $[R(\theta)w]$ is invertible. Therefore, the quotient matrix $c_i$ and $[m_i]^2$ share the same eigenvalues. While the transmittance amplitudes can be calculated from the two real eigenvalues $\lambda_1$ and $\lambda_2$, as $t_X=\sqrt[4]{\lambda_1}$ and $t_Y=\sqrt[4]{\lambda_2}$, the retardance of the calibration sample $\phi$ is a function of the two complex eigenvalues $\lambda_3$ and $\lambda_4$, as $\phi=\left|{\rm arg}(\lambda_3)-{\rm arg}(\lambda_4)\right|$.

Error sources depending on the measurement time scale are categorized into stochastic noise and systematic errors. Characteristic time scales are the total measurement time for intensities without the calibration samples $i_0$, those with the calibration samples $i_i$ and the sample-switch time in between. Stochastic noise with a time scale shorter than the total measurement time, comes from the laser source, the CCD, vibration of the rotatory positioners and the mechanical mounts of the optical elements. Each pixel of the CCD has a fluctuation of $\pm 0.3\%$ in the measured intensity which is modeled as statistically the same, and comes primarily from the repeatability of the laser source \cite{DUV2003} and the random spatial non-uniformity in the CCD \cite{EMVA1288}. Both cross-talk between neighboring pixels and electrical shot noise contribute to the spatial non-uniformity. Cross-talk is simulated via the correlation length of these noise sources across the CCD. The correlation length is taken to be 1 pixel for simplicity, i.e. no cross-talk is assumed. For longer correlation lengths, filtering algorithms may be applied to reduce the noise influence. The impact of electrical shot noise on the signal to noise ratio decreases as the number of photons increases (assuming the photon-to-electron conversion rate of 1). By carefully selecting the measurement conditions so that the CCD is near saturation, electrical shot noise buried in the signal controlled by the power of the laser and the integration time of the CCD can have less than 1/10 of the influence on diattenuation and retardance caused by the quantization noise due to the analogue-to-digital conversion (ADC) of the CCD. Electrical shot noise can therefore be safely neglected under the assumption of near CCD saturation, 10$^{14}$ photons per pixel in the model. The stochastic vibration of the rotatory positioners attached to {polarization elements} P1, Q1, Q2, and P2 in axial direction is taken to be 0.01\D. This follows from the Thorlabs’ motorized rotator K10CR1 \cite{ThorlabsMotor} specifications. Tilted variation of the PSG and PSA on the other hand is allocated to the polarization properties of the mirror and the BS in addition to the stochastic noise of the retardance and reflectance across the mirror.
The pre-measurement of the mirror can be performed by a single pass polarimeter in reflection mode using analysis \cite{Compain99} similar to this step to obtain those 
{stochastic noise}. The difference lies in that for the double-pass layout the light probes a SUT in both forward and backward directions, while in the single pass polarimeter the light incidents on a SUT (mirror here) only once. To calibrate the mirror under normal incidence, an additional BS is required to deflect the beam from reflection in the single pass polarimeter and should be calibrated in advance. The tolerance types for stochastic noise and their values are summarized in Tab.~\ref{Tab:stoch-noise}.
\begin{table}[h!]
\caption{Tolerance types and their values for stochastic noise}
      \begin{tabular}{ccccc}
        \hline
         Tolerance type & value & source & in step \\ \hline
        $\Delta I_0^{x,y},\Delta I_i^{x,y} $ & 0.3\% & laser and CCD & 1,2,3 \\
        $\Delta R_X,\Delta R_Y $ & 0.0002 & mirror & 1,2 \\
        $\Delta R_X,\Delta R_Y$ & 0.0002 & hemispherical mirror & 3 \\
        $\Delta T_X ,\Delta T_Y$ & 0.0002 & calibration samples & 2 \\
        $\Delta \Phi $ & 0.03\D & mirror & 1,2\\ 
        $\Delta \Phi $ & 0.03\D & hemispherical mirror & 3\\ 
        $\Delta \Phi $ & 0.012\D & calibration samples & 2\\ 
        $\Delta \Theta $ & 0.01\D  & P1, Q1, Q2, P2, and BS & 1,3\\
         $\Delta \Theta $ & 0.01\D  & mirror & 1,2\\
        $\Delta \Theta $ & 0.2\D\ \cite{ThorlabsMotor} &\qquad hemispherical mirror  & 3 \\ 
        $\Delta \Theta $ & 0.2\D\ \cite{ThorlabsMotor} &\qquad calibration samples  & 2 \\\hline
      \end{tabular}
      \label{Tab:stoch-noise}
\end{table}

Elements of the measured intensity matrix in Eq.~\eqref{i0matrix} equal to the true values plus errors, $i_0^{x,y}=I_0^{x,y}+\Delta I_0^{x,y} ({\rm where\ }x=1,2,...,u; y=1,2,...,v)$. The measured azimuthal angle $\theta = \Theta + \Delta\Theta$ in Eq.~\eqref{ci} is the sum of true value $\Theta$ and precision $\Delta\Theta$ of rotatory positioners. The reflectance error of the mirror $\Delta R_{X/Y} = r_{X/Y} - R_{X/Y} $, the retardance error $\Delta \Phi=\phi-\Phi$ and the transmittance amplitude error $\Delta T_{X/Y}=t_{X/Y}-T_{X/Y}$ all follow the same convention.

Although the PSG and PSA nominally have systematic errors, with the settings of the PSG and PSA being the same between the calibration (Fig.~\ref{Fig:layout}a) and measurement (Fig.~\ref{Fig:layout}b) setups there is no systematic change in the errors for the PSG and PSA matrices $w$ and $a$. Therefore, the systematic error introduced in the modeling comes from the mirror in steps 1-2, the objective and the hemispherical mirror in step 3. The CCD is a common element in both the calibration and measurement layout, nevertheless, information loss in the process of ADC cannot be calibrated out. Hence the systematic error from the CCD must be included in all 3 steps. Systematic errors are listed in Tab.~\ref{Tab:sys-noise}.

\begin{table*}[!hptb]
	\caption{Tolerance types and their values for systematic errors}
	\begin{tabular}{ccccc}
		\hline
		Tolerance type  & value  & source & in step\\ \hline
		$\Delta I_0^{x,y},\Delta I_i^{x,y}$ & 10 bit & CCD & 1,2,3 \\
		$\Delta R_X$ & $1-\sqrt{0.45}$ & mirror & 1,2\\
		$\Delta R_Y$ & $1-\sqrt{0.43}$ & mirror & 1,2\\
		$\Delta R_X$ & $1-\sqrt{0.5}$ & hemispherical mirror & 3 \\
		$\Delta R_Y$ & $1-\sqrt{0.4}$ & hemispherical mirror & 3\\
		$\Delta \Phi$ & 3\D & mirror & 1,2 \\
		$\Delta \Phi$ & 3\D & hemispherical mirror & 3  \\ 
		$\Delta \Theta$ & 0.2\D  & mirror  & 1,2 \\ 
		Diattenuation & 0.115\%  RMS OZP & objective in forward path & 3\\
		Retardance & 0.15\D\  RMS OZP & objective in forward path & 3\\
		Diattenuation & 0.114\% RMS OZP & objective in backward path & 3\\
		Retardance & 0.14\D\ RMS OZP & objective in backward path & 3 \\\hline	
	\end{tabular}
	\label{Tab:sys-noise}
\end{table*}

\begin{figure}[h!]
\begin{center}
\includegraphics[width=0.76 \columnwidth]{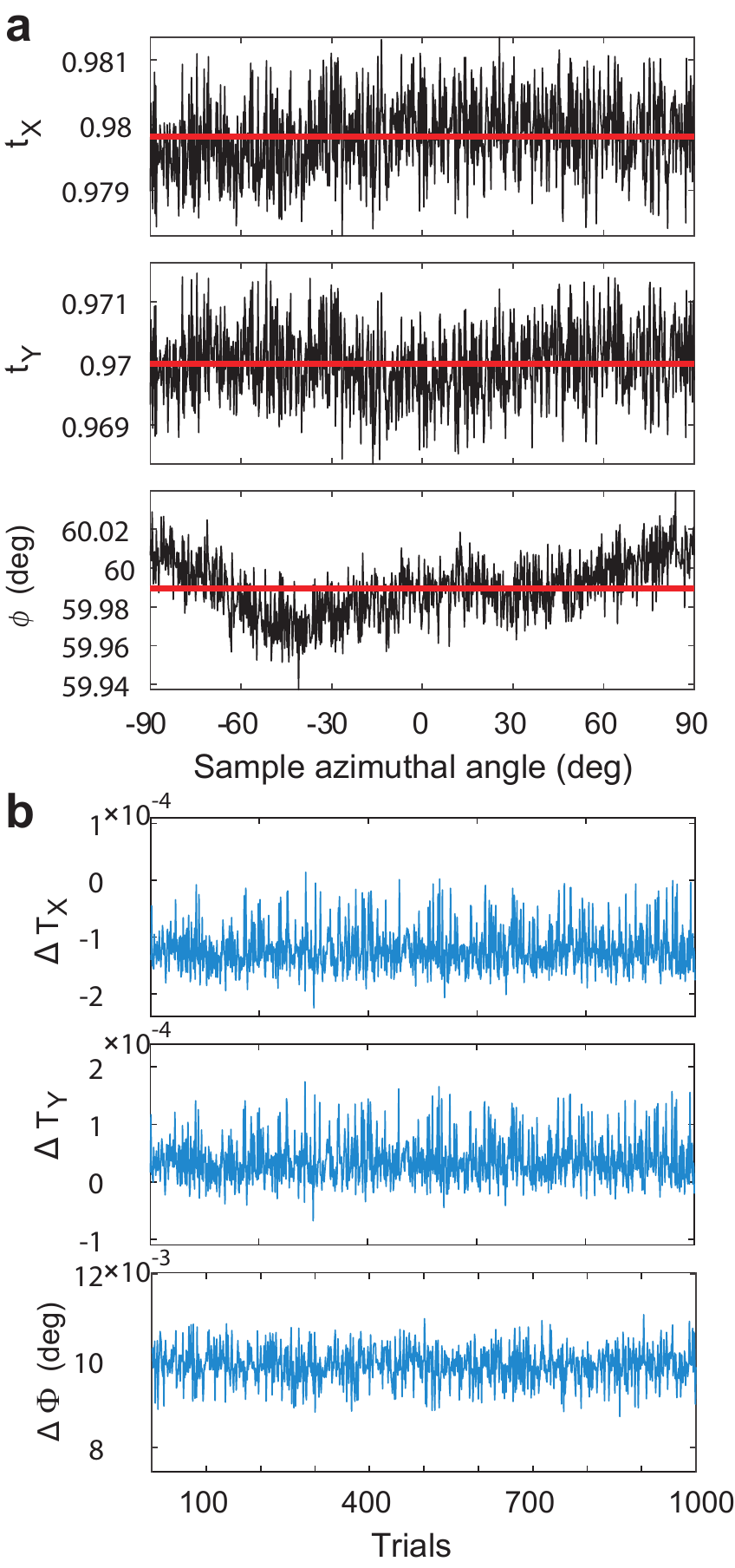}
\end{center}
  \caption{The measurement results of and stochastic noise in the polarization properties of the 1/6-wave plate in a double-pass setup. {\bf a} Measurement results of the transmittance amplitude $t_X$, $t_Y$, and the retardance $\phi$ as a function of the azimuthal angles from the -90\D\ to 90\D\ range. The red line is the average measurement value over all measured azimuthal angles. The true transmittance amplitudes are $T_X=0.98$, $T_Y=0.97$, and the true retardance is $\Phi = 60$\D.
  {\bf b} The stochastic noise for 1000 trials. For each trial the 
stochastic noise is defined as the average measurement (red line) minus the true value of each polarization property.}
\label{Fig:DP-waveplate}
\end{figure}

With the stochastic and systematic errors 
{of each component of the polarimeter listed above, we simulate both the stochastic noise and systematic errors
in the properties of the calibration samples using a bottom-up approach.} To reduce the rotational asymmetrical noise such as the tilt angle of the calibration samples, we rotate the calibration sample azimuthally and take the average over the -90\D\ to 90\D\ range. Figure ~\ref{Fig:DP-waveplate}a shows the calibrated transmittance amplitudes $t_X$, $t_Y$ and retardance $\phi$ as a function of the azimuthal angle of the 1/6-wave plate sample. The average measurement value is displayed as a red line with the true values of the transmittance amplitudes being $T_X = 0.98$, $T_Y=0.97$ and that of the true retardance being $\Phi = 60$\D. The 
stochastic noise is defined as the difference between the average measurement over all azimuths and the true value. A 1000-trial simulation in Fig.~\ref{Fig:DP-waveplate}b indicates that the calibration stochastic noise for transmittance {is} $\Delta T_X<\pm0.0002,\ \Delta T_Y <\pm0.0002$ and that for retardance is $\Delta \Phi <\pm 0.012$\D\ as shown in Tab.~\ref{Tab:stoch-noise}. For the calibration sample polarizer, this step is sufficient to determine the transmittance amplitude of the bright transmission axis, but not the dark axis. Using two 40 dB polarizers in series could ensure a stochastic error $\Delta T_Y <\pm0.0002$. The calibrated polarization properties of the 1/6-wave plate and polarizer, together with their stochastic noise determines the calibration accuracy for the PSG $w$ and PSA $a$ matrices in step 2.

\subsection{Step 2: Calibration of the PSG and PSA matrices}
\begin{figure}[h!]
\begin{center}
\includegraphics[width=0.63 \columnwidth]{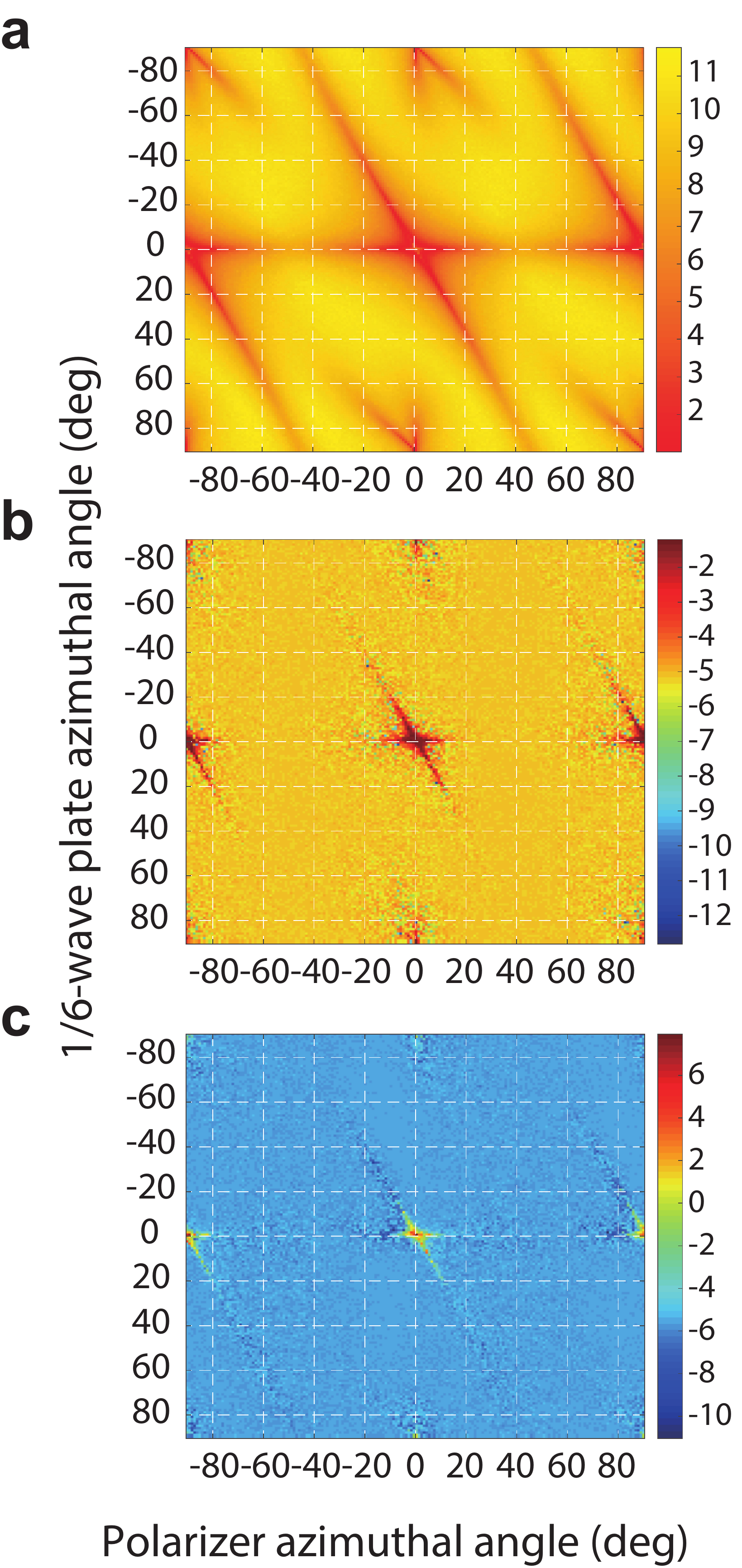}
\end{center}
  \caption{ Maps of the error estimator, the 
{calibration} error matrix for the PSG $\Delta W$ and the PSA $\Delta A$ as a function of the azimuthal angles of the polarizer and the 1/6-wave plate. {\bf a} The error estimator is taken to be log[$\lambda(2)/\lambda(1)$], where $\lambda(2)/\lambda(1)$ is the quotient of the two smallest eigenvalues for $k$ in Eq.~\eqref{eqk}. {\bf b} The error between the true and measured values of the PSG matrix in logarithm log($\Delta W$). The element (1,1) in the 4$\times$4 matrix is presented. {\bf c} The error of the (1,1) element in the PSA matrix in logarithm log($\Delta A$). The measured and true matrices for the PSG and PSA are normalized by their transmittance.}
\label{Fig:logR-angles}
\end{figure}

In this subsection, we calculate the PSG $w$ and PSA $a$ matrices as well as expand the working range of the azimuthal angles of the calibration samples from those used in the past \cite{Laude2004}. To compute the PSG matrix $w$, $w$ in Eq.~\eqref{ci} is first replaced with an unknown matrix $x$. This results in $m_i^{\rm DP}x-xc_i=0$,  where the sample matrix for the double-pass polarimeter is defined as $m_i^{\rm DP}\equiv m^{-1}_{\rm mirror}m^b_im_{\rm mirror}m^f_i$. A linear operator of the $i$\textsuperscript{th} calibration sample $h_i$ is related to the quotient matrix $c_i$ and the sample matrix $m_i^{\rm DP}$ by
\begin{eqnarray}
h_i(m_i^{\rm DP},c_i)x=0.
\label{hi}
\end{eqnarray}
The linear operator $h_i$ is a 16$\times$16 matrix, the elements of which are detailed in Eqs.~\eqref{hiexpa} and \eqref{GaG} in Appendix~A. The calibrated PSG matrix $w$ is then the non-zero solution to Eq.~\eqref{hi}. The Mueller matrix of the calibration samples in the double-pass $m_i^{\rm DP}$ including the transmittance amplitudes $t_X$, $t_Y$, the retardance $\phi$  of $m_i^b$ and $m_i^f$ are calculated from step 1, while the azimuthal angle error $\Delta \Theta $ in the rotation matrix $R(\theta=\Theta+\Delta \Theta)$ in Eq.~\eqref{ci} is determined by the commercial rotary positioners repeatability of 0.2\D\ as listed in Tab.~\ref{Tab:stoch-noise}. A Hermitian $k$ matrix is defined as the transpose of the linear operator $h_i$ times itself with the summation of all calibration samples
\begin{equation}
k(m_i^{\rm DP},c_i)x=(\sum_i h_i^Th_i)x .
\label{eqk}
\end{equation}  
In this way, all 16 calculated eigenvalues $\lambda(1)<\lambda(2)<\hdots<\lambda(16)$ of $k$ must be positive and real. The eigenvector with eigenvalue closest to 0 is the calculated PSG $w$ matrix, after the 16$\times$1 eigenvector {being reshaped} into a $4\times 4$ matrix. Three calibration samples $i=3$ are enough \cite{Stabo2009}, and the first sample is a polarizer with azimuthal angle $\Theta=0$\D. To find suitable combinations of azimuthal angles that guarantee calibration accuracy, we plot the error estimator log[$\lambda(2)/\lambda(1)$] as a function of the azimuthal angles for calibration sample 2 (a polarizer with different azimuthal angle to sample 1) and sample 3 (a 1/6-wave plate) in Fig.~\ref{Fig:logR-angles}a. Both of their azimuthal angles are varied from -90\D\ to 90\D.

Stochastic noise contributions to the error estimator include the rotational repeatability of the calibration samples as limited by mechanical positioners, intensity fluctuations, stochastic polarization noise of the calibration samples in step 1 and that of the mirror. Systematic errors come from the quantization error of the ADC, and the polarization properties of the mirror.  Values of these errors are given in Tabs.~\ref{Tab:stoch-noise} and \ref{Tab:sys-noise}. Each combination of azimuthal angles in Fig.~\ref{Fig:logR-angles}a is averaged over 100 trails to reduce the influence from stochastic noise. The larger the value of the error estimator, the closer the smallest eigenvalue of $k$ in Eq.~\eqref{eqk} is to 0, and consequently the more accurate the calculated PSG $w$ matrix will be. We observe that the error estimator is relatively small, log[$\lambda(2)/\lambda(1)] < 9$, when the 1/6-wave plate has the same azimuthal angle ($\Theta\approx 0$\D, the middle horizontal reddish line in Fig.~\ref{Fig:logR-angles}a) as that of the first polarizer. It is likely that the lack of calibration accuracy is due to the azimuthal angle overlap of the two orthogonal eigenstates of the first polarizer and the 1/6-wave plate, blurring the precision of the eigenvalue-based calibration method.

We further calculated the calibration error between the calibrated and the true PSG matrices $\Delta W $ and that of the PSA error matrix $\Delta A $. The calculated PSG $w$ matrix is normalized by its transmission before the comparison with the true PSG matrix $W$, because the eigenvector of Eq.~\eqref{eqk} can be scaled with any real number. As the calculated PSA matrix $a$ is derived from the measured intensity using Eq.~\eqref{i0matrix}, it will give an inverse scaling factor to the calculated PSG matrix $w$ if the normalization is not done. Consequently, normalization of transmission only serves for obtaining the error for the PSG $\Delta W$ and the PSA $\Delta A$. The PSG $w$ and PSA $a$ matrices without normalization will not affect the measurement accuracy of a SUT in step 3. The logarithm of the error of the PSG $\Delta W$ and the PSA $\Delta A$ as a function of the azimuthal angles of the 1/6-wave plate and the polarizer are plotted in Fig.~\ref{Fig:logR-angles}b and Fig.~\ref{Fig:logR-angles}c, respectively. The first element (1,1) of the 4$\times$4 error matrices can be chosen without loss of generality. The other 15 elements of the error matrices $\Delta W$ and $\Delta A$ share roughly the same calibration error. The cross areas in the middle of the error matrices for the PSG and PSA display a relatively worse accuracy, and are aligned with the error estimator map, log[$\lambda(2)/\lambda(1)$] in Fig.~\ref{Fig:logR-angles}a. As a result, the requirement for alignment of the calibration samples can be relaxed to all the yellowish areas in Fig.~\ref{Fig:logR-angles}a, corresponding to the error estimator log[$\lambda(2)/\lambda(1)$] $>$ 10.

Former experimental observations reveal the calibration accuracy of the PSG and PSA matrices, where an average of a standard deviation over all 16 Mueller matrix elements is employed for quantification \cite{Stabo2009}. In those experiments, the averaged standard deviation is 6.7$\times 10^{-4}$ for the PSG matrix and 6.0$\times 10^{-4}$ for the PSA matrix over 38 calibrations. We simulate the pixel-based PSG and PSA matrices for 10000 trails, and obtain the averaged standard deviation of 5.9$\times 10^{-4}$ for the calibrated PSG and that of 3.6$\times 10^{-4}$ for the calibrated PSA, which is in line with the experiments, verifying our tolerance analysis for the calibration. 

\subsection{Step 3: Angle-resolved measurement}
The alignment of the objective to the center of the hemispherical mirror can be monitored by adding an interferometer arm to form an interference pattern on the CCD. This added arm would extend horizontally from the laser and the PSG, and have a mirror at the end.

\begin{figure}[h!]
\begin{center}
\includegraphics[width=1 \columnwidth]{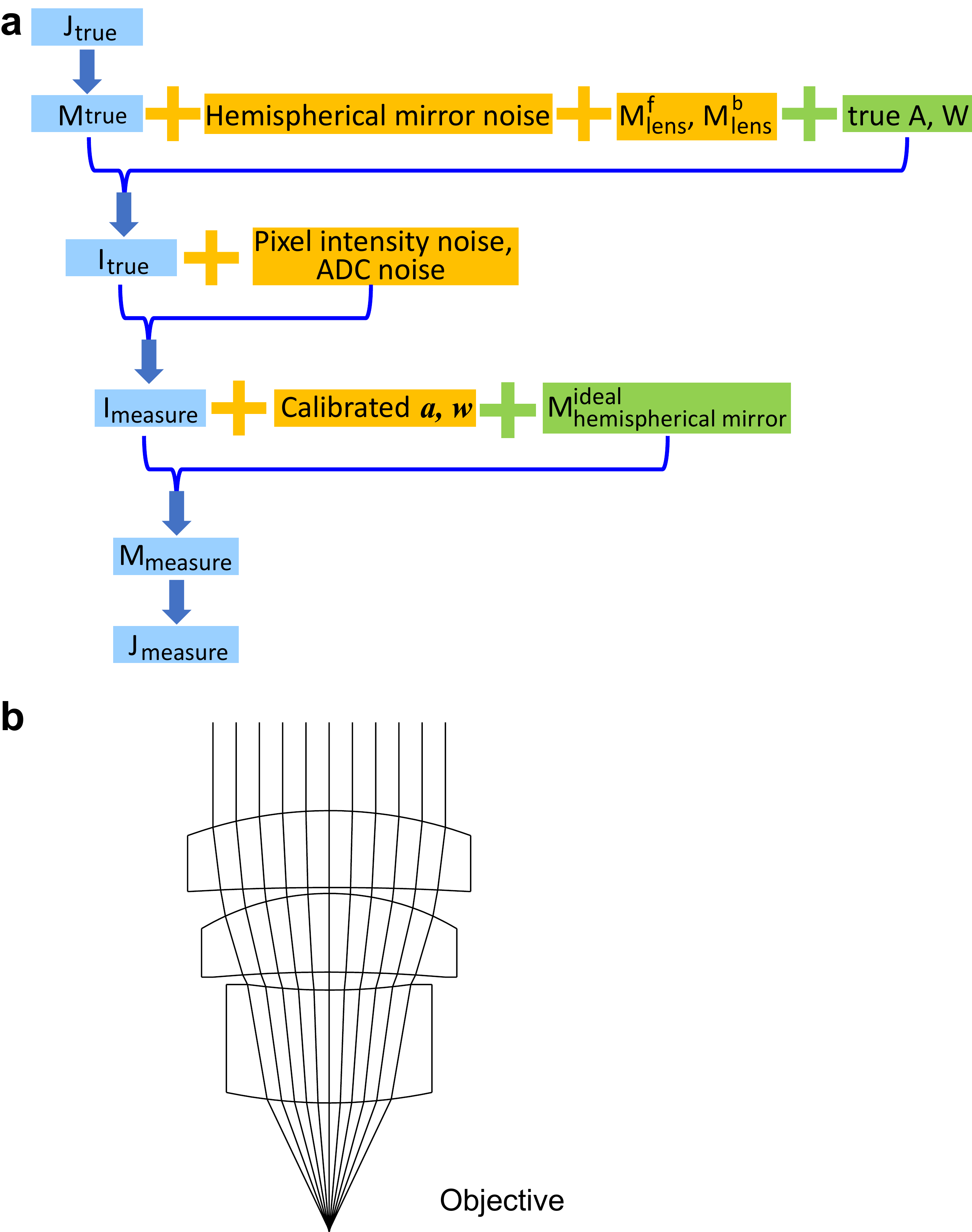}
\end{center}
  \caption{The simulation flow for prediction of the polarization sensitivities and an objective layout. {\bf a} Blue blocks: parameters being tracked through the simulation; yellow blocks: stochastic noise and systematic errors; green blocks: idealized or true values of optical elements. {\bf b} The layout of the objective used in step 3 with an incident angle of 25.4\D\ from a Japanese patent 61$\_$2925 860129 in the Code V patent database \cite{CodeV}.  }
\label{Fig:simu_flow_lens}
\end{figure}

The simulation flow leading to the prediction of the sensitivities for the angle-resolved measurements is depicted in Fig.~\ref{Fig:simu_flow_lens}a. The simulation uses a generated Jones matrix covering the whole pupil (in short Jones pupil matrix) as the true Jones pupil matrix of a SUT $J_{\rm true}$. It is synthesized by the RMS of the coefficients of up to order 72 in an expansion using the OZP for diattenuation and retardance \cite{Tilmann2008,Ruoff2009,Ruoff2010}.

The Jones pupil matrix is converted to a Mueller pupil matrix to be compatible with the Mueller matrix description $M_{\rm true}$ of the PSG and PSA in step 1-2. The true PSG matrix $W$, PSA matrix $A$, stochastic noise from the hemispherical mirror (whose values are listed in Tab.~\ref{Tab:stoch-noise}), combined with the systematic errors from the objective and hemispherical mirror  (whose values are listed in Tab.~\ref{Tab:sys-noise}), result in the true intensity $I_{\rm true}$. Objectives usually contain multiple lenses to ensure a specific image quality over the field of view. As polarization relies on the order of the components the light passes through, the polarization of the light traveling through the objective in the forward direction, from collimated space to the focus as sketched in Fig.~\ref{Fig:simu_flow_lens}b from top to bottom, differs from the light transmitted by the objective in the backward or return direction. As an example, we choose a Japanese patent 61\_ 2925 860129 in the CODE V database \cite{CodeV} with a half incident angle of 25.4\D to investigate its polarization properties. Without applying optical coatings to the objective, we trace the polarization of the objective in both the forward and backward directions.
{The backward beams exiting the object have maximum deviation angles of 0.29\D\ along the periphery due to the imperfect wavefront of the objective. Systematic errors of diattenuation and retardance of the objective considering the retrace error in the backward direction are listed in Tab.~\ref{Tab:sys-noise}.}
Random intensity noise at each pixel and information loss from the ADC are added to the intensity as error sources to form the measurement intensity $I_{\rm measure}$. The calibrated PSG matrix $w$, PSA matrix $a$, and idealized Mueller matrix [1 0 0 0; 0 1 0 0; 0 0 -1 0;0 0 0 -1] are employed to calculate the Mueller matrix of the SUT in the forward path $M_{\rm measure}$. The Mueller pupil matrix is converted to the Jones pupil matrix afterwards. This procedure removes the information about depolarization contained in the Mueller matrix to obtain the Jones matrix. Depolarization in the measured Mueller matrix $M_{\rm measure}$ comes from overlap of incoherent electromagnetic fields \cite{advanced2013}. To convert the Mueller matrix with limited depolarization to the Jones matrix $J_{\rm measure}$, the non-depolarization condition for the conversion ${\rm trace}(M^TM)=4m_{11}^2$ \cite{Simon1987} is approximated as $|{\rm trace}(M^TM)-4m_{11}^2|<0.01$. For measurements that meet this condition, the Jones matrix can be derived from the Mueller matrix via expressions given in Ref.~\cite{Savenkov1997}.

The RMS of the OZP coefficients for either diattenuation or retardance is a single number used to quantify the goodness of a Jones pupil via the relative transmittance amplitude difference between the brightest and darkest axes or retardance delay between the fastest and slowest axes across the pupil of a SUT, respectively. Mathematical details of the OZP can be found in Appendix B. For 72 terms the highest power in the radial direction of the OZP is 10, corresponding to the highest radial power of the 36\textsuperscript{th} term of the fringe Zernike polynomials \cite{CodeV}.

Though the true reflectance amplitude of the hemispherical mirror is not unity, only the difference between the reflectance in the X and Y directions affects the diattenuation and retardance of the pupil. This is because the measured Jones pupil matrix $J_{\rm measure}$ is further decomposed into a product of apodization, a partial polarizer, a retarder and two other physically meaningful matrices \cite{Geh2007}, and only the diattenuation pupil in the partial polarizer and the retardance pupil in the retarder will be further expanded by the OZP. Writing the reflectance in the X and Y directions of the hemispherical mirror as $r_X=r_Y+\Delta R_{\rm XY}$, the average of the reflectance in the X and Y direction contributes only to the apodization of the SUT. The difference of the reflectance amplitudes $\Delta R_{XY}$ will be counted in the first term of the OZP expansion (see Eqs.~\eqref{Jpoloz} and \eqref{Jretoz} for the mathematics). Since rotating the hemispherical mirror azimuthally for 90\D\ swaps the reflectance values $r_X$ and $r_Y$, taking the average of the fitting coefficients to the OZP expansion, measured with 0\D\ and 90\D\ hemispherical mirror rotation, improves the accuracy of the OZP coefficients for the diattenuation and retardance pupils.

We decompose both $J_{\rm true}$ and $J_{\rm measure}$ into an OZP description of retardance and diattenuation, using the first 72 terms. The RMS of the coefficients are calculated as RMS $= \sqrt{\sum_{j=1}^{72} coe_j/(j+1)}$, with $coe_j$ denoting the $j$\textsuperscript{th} OZP coefficient. Comparison of the true Jones pupil matrix for the SUT and the measured value is made by running the simulation through the flow in Fig.~\ref{Fig:simu_flow_lens}a for 100 trials. The repeatability in terms of the RMS of the OZP coefficients replaces the mean value in the standard variance \cite{iso1} with the true value, defined as
\begin{align}
\begin{split}
& {\rm repeatability} \equiv \\
& \sqrt[4]{\frac{\sum_1^{\rm trail}|{\rm RMS\_OZP^2_{meas,trail}}-{\rm RMS\_OZP_{true}^2}|^2}{\rm trail -1}}.
\end{split}
\end{align}

Before predicting the sensitivity presented by RMS OZP in the Jones matrix description, we apply our tolerance analysis to the SUT in terms of the Mueller matrix in the measurement procedure similar to that reported experimental observations in Ref.~\cite{Compain99}. We compare the true Mueller matrix $M_{\rm true}$ with the measured Mueller matrix $M_{\rm measure}$ in the simulation flow as sketched in Fig.~\ref{Fig:simu_flow_lens}a. Both of the two matrices are normalized to their (1, 1) elements, so that the relative error of the (2, 2), (3, 3) and (4, 4) elements of the matrices can be calculated under the condition of a non-identity Mueller matrix of the mirror. Off-diagonal elements of the Mueller matrices $M_{\rm true}$ and $M_{\rm measure}$ are small due to the generated weak polarization properties of the SUT, leading to unphysically large relative errors, and thus they are safely disregarded in the comparison. We obtain a maximum 0.4\% over all three Mueller matrix pupils, in good agreement with the 0.5\% in the reported experiment.    

\section{Results and discussion}
\begin{figure}[h!]
\begin{center}
\includegraphics[width=0.72 \columnwidth]{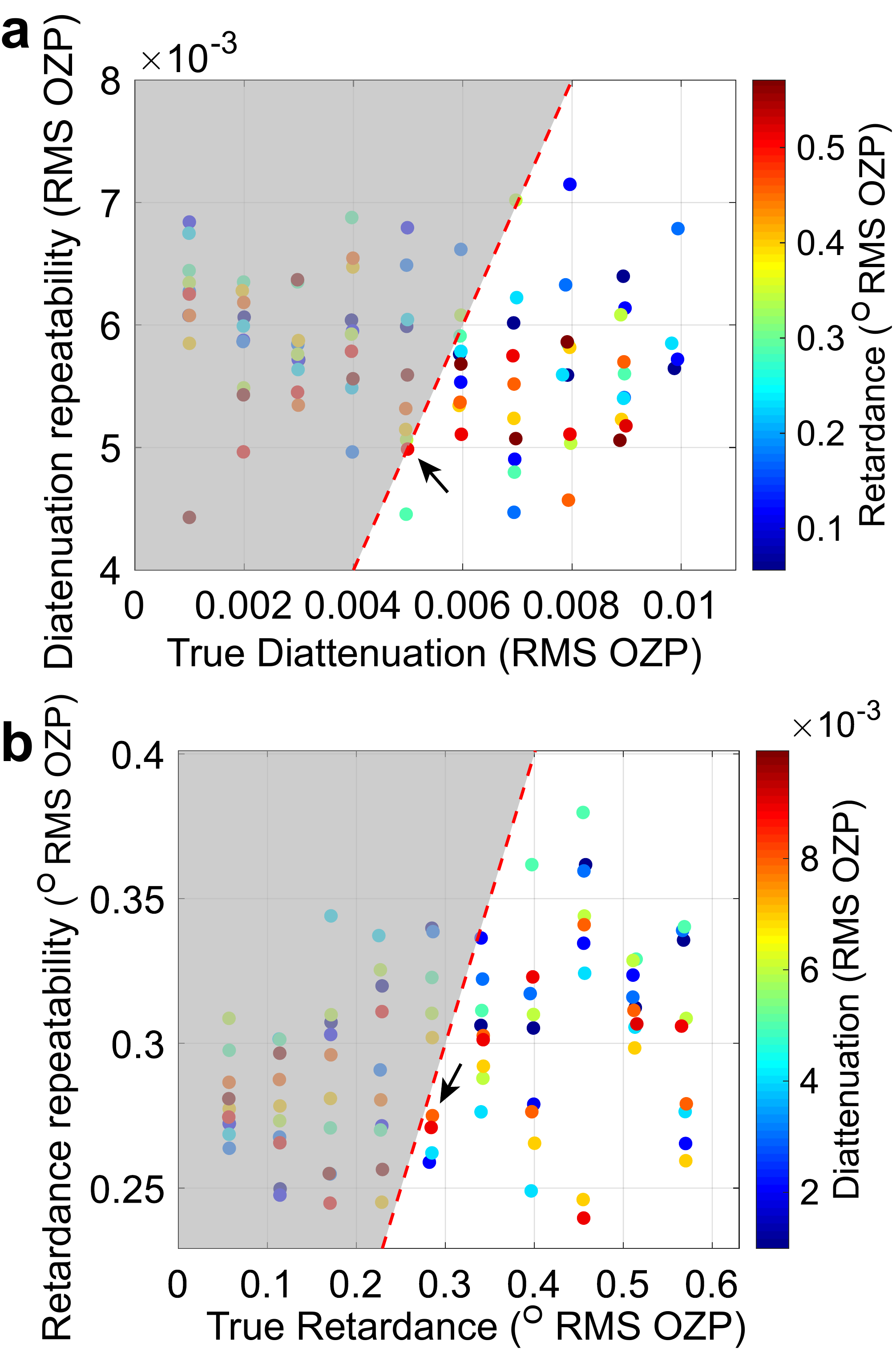}
\end{center}
  \caption{Sensitivities of diattenuation and retardance. {\bf a} Repeatability of diattenuation as a function of its true value expressed in RMS OZP. Colors denote the true values of the corresponding retardance of the pupils. A dotted red line going through the origin (0,0) has a slope of 1, where the true values equal to repeatability, defining sensitivity. Pupils with the polarization properties inside gray shaded areas have worse repeatability than the true values. {\bf b} Repeatability of retardance versus the true retardance of the pupils. Colors denote the true values of the corresponding diattenuation of the pupils expressed in RMS OZP.  Each repeatability is calculated from 100 trials, where each trial follows the complete flow in Fig.~\ref{Fig:simu_flow_lens}a. {Black arrows point to the pupils that meet our definition of sensitivity, and polarization properties of these pupils are visualized in Fig.~\ref{Fig:decomp_meas}.}}
      \label{Fig:100_dia_ret}
      \end{figure}

Sensitivity is defined in terms of a boundary. In Fig.~\ref{Fig:100_dia_ret}, the boundary where repeatability equals the true value is the line with a slope of 1 through the origin (0,0). Away from the gray shadow areas, the repeatability (i.e. the measurement uncertainties) are smaller than the true values. The sensitivity of the diattenuation pupil depends on the corresponding retardance. Larger retardance leads to better sensitivity for diattenuation in general. The same phenomenon applies to the sensitivity of retardance as well. It is likely that the measurement is more sensitive when the SUT exhibits strong polarization properties, and the retardance and diattenuation are not decoupled in calculating the repeatability of either of them. To reduce the sensitivity from a set of values to a single value, we quantify the sensitivity of diattenuation with an additional requirement: the corresponding retardance of the pupil 
should be of the same order of magnitude as the diattenuation. This results a 0.5\%  RMS OZP sensitivity for diattenuation. With the same requirement, the predicted sensitivity for retardance is 0.3\D\ RMS OZP.

\begin{figure}[h!]
\begin{center}
\includegraphics[width=0.96 \columnwidth]{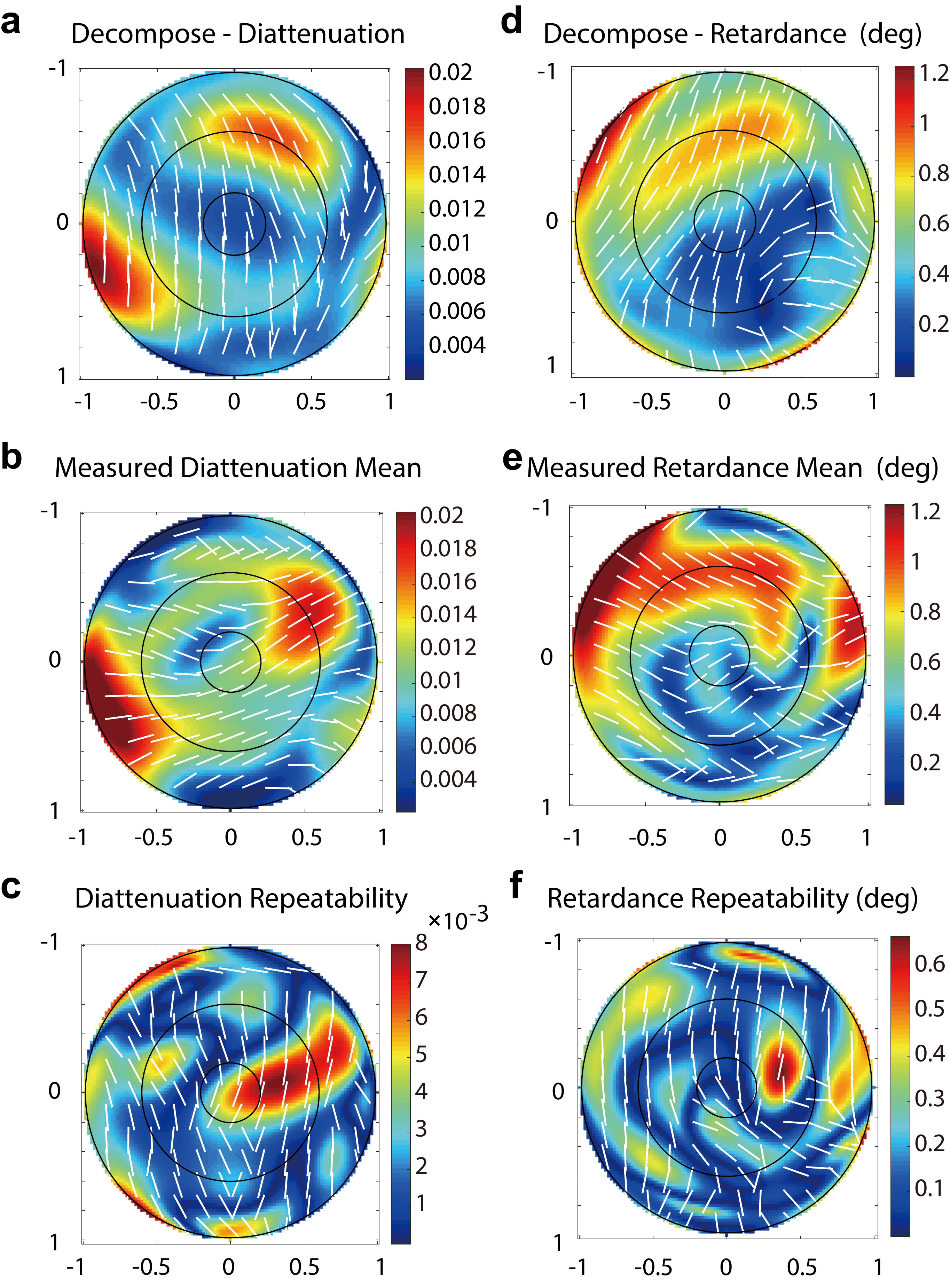}
\end{center}
  \caption{Visualization of the pupils for diattenuation and retardance when their repeatabilities equal to their true values. {\bf a} A SUT pupil of true diattenuation decomposed from a pupil of the Jones matrix. The Jones matrix pupil has diattenuation of 0.5\% RMS OZP and retardance of 0.5\D\ RMS OZP. {\bf b} The mean of the measured pupil of diattenuation and {\bf c} the repeatability pupil of diattenuation reconstructed from the first 72 terms of the OZP expansion over 100 trials. {\bf d} A pupil of true diattenuation after the Jones matrix pupil decomposition. The Jones pupil has diattenuation of 0.8\% RMS OZP and retardance of 0.3\D\ RMS OZP. {\bf e} The mean of the measured pupil of retardance and {\bf f} the repeatability pupil of retardance from reconstruction over 100 trials. From inside to outside,  concentric circles correspond to angles at the objective of 5\D, 15\D and 25\D. White directional lines denote the azimuthal angles across the pupils.}
      \label{Fig:decomp_meas}
      \end{figure}

Visualization of the pupils for {the true} diattenuation and retardance when their sensitivities are reached
{(labeled with black arrows in Fig.~\ref{Fig:100_dia_ret}),}
i.e. 0.5\% RMS OZP for diattenuation and 0.3\D\ RMS OZP for retardance, is shown in Fig.~\ref{Fig:decomp_meas}a and Fig.~\ref{Fig:decomp_meas}d, respectively. The mean of the measured pupils of diattenuation and retardance comes from the measured Jones pupil $J_{\rm measure}$, and a decomposition of the measured Jones pupil in terms of diattenuation and retardance thereafter. Reconstruction of the pupils for diattenuation and retardance is based on the first 72 terms of the OZP expansion, where each pixel of the pupils for diattenuation and retardance is averaged over 100 trials. The sensitivity of diattenuation shows an average of 1\% over {all pixels of} a pupil displayed in Fig.~\ref{Fig:decomp_meas}b with repeatability around 1/3 of that displayed in Fig.~\ref{Fig:decomp_meas}c. For the sensitivity of retardance, the average of the pupil is 0.6\D\ as  shown in Fig.~\ref{Fig:decomp_meas}d with a repeatability around 1/3 of that as well, as shown in Fig.~\ref{Fig:decomp_meas}f. Directional lines on the diattenuation pupils denote azimuthal angles for the partial polarizer, while they denote those for the retarder on the pupils of retardance. The azimuthal angle pupil reconstructed from the OZP coefficients may have a 90\D\ shift, due to the limitation of the inverse trigonometric functions described in Eqs.~\eqref{gamma} and \eqref{beta} in Appendix B. Horizontal lines represent an azimuthal angle of 0\D , while vertical lines represent 90\D . White lines with directions other than vertical or horizontal represent error, the larger the error of the direction, the farther away the direction of the white line is from either vertical or horizontal.
When treating the 90\D\ shift to be error-free, the repeatability of the azimuthal angle is 5\D\ for diattenuation and 3\D\ for retardance averaged across the pupil, lower than 1/10 of the mean values of that for  diattenuation and retardance.
   
Concentric circles with different radii on the pupil correspond to different incident angles of the laser beam away from the focal plane of the objective. In Fig.~\ref{Fig:decomp_meas}, from inside to outside the concentric circles correspond to angles at the objective of 5\D, 15\D and 25\D. Pupils of diattenuation and retardance provide a visualization of the azimuthal anisotropy and polarization response of a refractive sample under non-normal incidence. 

\section{Conclusion}
In conclusion, we have performed a detailed tolerance analysis of the calibration and measurement procedures for a double-pass polarimeter, and have predicted the sensitivity of the polarimeter to systematic errors and stochastic noise. The eigenvalue calibration method ECM \cite{Compain99} is used in the polarimeter calibration, resulting in the Mueller and Stokes description of the PSG and PSA characteristic matrices. The Mueller pupil matrix of 
{an arbitrary} non-depolarizing SUT is predicted before it is converted to a Jones pupil matrix.
{Our tolerance model for the calibration of the PSG and PSA, as well as the measurement of the Mueller matrix pupil are consistent with previous experimental observations \cite{Compain99,Stabo2009}. }
Thanks to the Jones pupil decomposition and the OZP expansions of diattenuation and retardance, the whole pupil of the SUT can be described by two values, diattenuation and retardance in terms of the RMS of the OZP coefficients. The sensitivity prediction for diattenuation is 0.5\% and that for retardance is 0.3\D. The double-pass polarimeter offers a platform to measure angle-resolved SUTs, revealing the azimuthal inhomogeneity of retardance and diattenuation. The ECM, tolerance analysis and the subsequent conversion of the measured Mueller pupil matrix of the SUT to a Jones pupil matrix in terms of the OZP expansions to predict sensitivities and visualize retardance and diattenuation pupils can also be applied to a single pass polarimeter. Though the incident angle would not be resolved in the single pass polarimeter, without a BS and a mirror fewer noise sources are included. The singe pass polarimeter can achieve better sensitivity of diattenuation and retardance as well as resolve the small inhomogeneity of the pupil under normal incidence.

\section{Acknowledgments}
The authors are thankful to Zejiang Meng for introducing the ECM algorithm, Vladimir Nikishkin for coding assistance, and Wei Wang for the initial contact with vendors for the specifications of the components used in the modeling as well as his exuberant personality. Last but not the least, we acknowledge an extended vacation due to COVID-19 outbreak.

\section*{}
\appendix
\section{Error propagation}
We have derived a simplified theory of error propagation for double-pass polarimetry to cross-check our numerical simulations with tolerances. By employing perturbation theory to the first order, we theoretically calculate the error of the PSG matrix $\Delta W$ given the measurables without any calibration samples for $i_0$ and with the calibration sample for $i_i$. 
\begin{eqnarray}
i_0 & = & am_{\rm mirror}w  \nonumber\\
& = & AM_{\rm mirror}W + \Delta ( AM_{\rm mirror}W)\\
i_i & = & am^b_{i}m_{\rm mirror}m^f_{i}w \nonumber\\
&  = & AM^b _{i}M_{\rm mirror}M^f_{i}W + \Delta ( AM^b_{i}M_{\rm mirror}M^f_{i}W ) \nonumber
\label{measure=true+error}
\end{eqnarray}

The noise propagation of the quotient matrix $c_i$ combining Eqs.~\eqref{measure=true+error} and \eqref{ci} to the first order results in the expression
\begin{eqnarray}
c_i & = &  W^{-1}M_i^{\rm DP}W  \nonumber\\
& & +\ W^{-1}M_{\rm mirror}^{-1}A^{-1}\Delta(AM_{\rm mirror}W)W^{-1}M_i^{\rm DP}W\nonumber\\
& & -\ W^{-1}M_{\rm mirror}^{-1}A^{-1}\Delta(AM_i^bM_{\rm mirror}M_i^fW).
\label{ciexpa}
\end{eqnarray}
Letting the unknown $x=W+\Delta W$, the linear operator $h_i(x)$ in Eq.~\eqref{hi} is expanded to the first order with the quotient matrix from Eq.~\eqref{ciexpa}, as
\begin{eqnarray}
& & h_i(W+\Delta W) \nonumber\\
&  = & M_i^{\rm DP}\Delta W -\Delta W W^{-1}M_i^{\rm DP}W\nonumber\\
&  + & \ \Delta(M_i^{\rm DP})W\nonumber\\
&  + & \ M_{\rm mirror}^{-1}A^{-1}\Delta(AM_{\rm mirror}W)W^{-1}M_i^{\rm DP}W\nonumber\\
&  - & \ M_{\rm mirror}^{-1}A^{-1}\Delta(AM_i^bM_{\rm mirror}M_i^fW).
\label{hiexpa}
\end{eqnarray} 

Factoring the first term on the RHS in Eq.~\eqref{hiexpa}, we operate on the elements of the matrices. It follows that by applying the relation for the least square fit $\Delta W_{p,q}=\delta_{p,F}\delta_{q,G}\Delta W_{F,G}$, where $\delta$ is the Kronecker delta, $p$, $q$, $F$ and $G$ are summed from 1 to 4, we have
\begin{eqnarray}
& & \Big [ M_i^{\rm DP}\Delta W  - \Delta W W^{-1}M_i^{\rm DP}W\Big ]_{p,q}\nonumber\\
& = & M_{i,p,o}^{\rm DP}\Delta W_{o,q} - \Delta W_{p,o}(W^{-1}M_i^{\rm DP}W)_{o,q}\nonumber\\
& = &	\Big [ M_{i,p,o}^{\rm DP} \delta_{o,F}\delta_{G,q} - (W^{-1}M_i^{\rm DP}W)_{o,q}\delta_{p,F}\delta_{o,G}\Big ]\Delta W_{F,G}\nonumber\\
& = & \Big [ M_{i,p,F}^{\rm DP} \delta_{G,q}-(W^{-1}M_i^{\rm DP}W)_{G,q}\delta_{p,F}\Big ]\Delta W_{F,G}\nonumber\\
& \equiv & G_{i,\mu,\nu}\Delta W_\nu.
\label{GaG}
\end{eqnarray}
The single indices $\mu$ and $\nu$ label all possible combinations of $F,G$ and $p,q$.
\begin{eqnarray}
& \mu=1  \rightarrow  p,q=1,1;  & \nu=1  \rightarrow  F,G=1,1 \nonumber\\
& \mu=2  \rightarrow p,q=1,2;  & \nu=2  \rightarrow  F,G=1,2 
\nonumber\\
& \vdots & \qquad\qquad \vdots 
\nonumber\\
& \mu=15  \rightarrow  p,q=4,3;  & \nu=15  \rightarrow  F,G=4,3 \nonumber\\
& \mu=16  \rightarrow  p,q=4,4; & \nu=16  \rightarrow  F,G=4,4 
\end{eqnarray}
The last two terms on the RHS in Eq.~\eqref{hiexpa} are influenced by the intensity with the calibration sample $\Delta(AM_i^bM_{\rm mirror}M_i^fW)$ and without it $\Delta(AM_{\rm mirror}W)$. Hence, the intensity error is defined as $\Delta I_{i,\mu}\equiv\Big [ M_{\rm mirror}^{-1}A^{-1}\Delta(AM_{\rm mirror}W)W^{-1}M_i^{\rm DP}W - M_{\rm mirror}^{-1}A^{-1}\Delta(AM_i^bM_{\rm mirror}M_i^fW) \Big]_{p,q}$. Assuming $G_{i,\mu,\nu}$ is invertible, Eq.~\eqref{hiexpa} can be simplified to
\begin{eqnarray}
\Delta W_\nu = - \Big ( G_{i,\mu,\nu} \Big )^{-1} \Big \{ \Big [\Delta(M_i^{\rm DP})W\Big ]_{i,\mu} + \Delta I_{i,\mu} \Big \}.
\label{hisim}
\end{eqnarray}
This expresses the linear relationship between one element of the PSG error matrix $\Delta W_{\nu}$ and the sum of the stochastic noise of the calculated calibration sample $\Delta (M_i^{\rm DP})$ times the true PSG matrix $W$ and the noise of the measured intensity $\Delta I_{i,\mu}$.

\begin{figure}[h!]
\begin{center}
\includegraphics[width=1 \columnwidth]{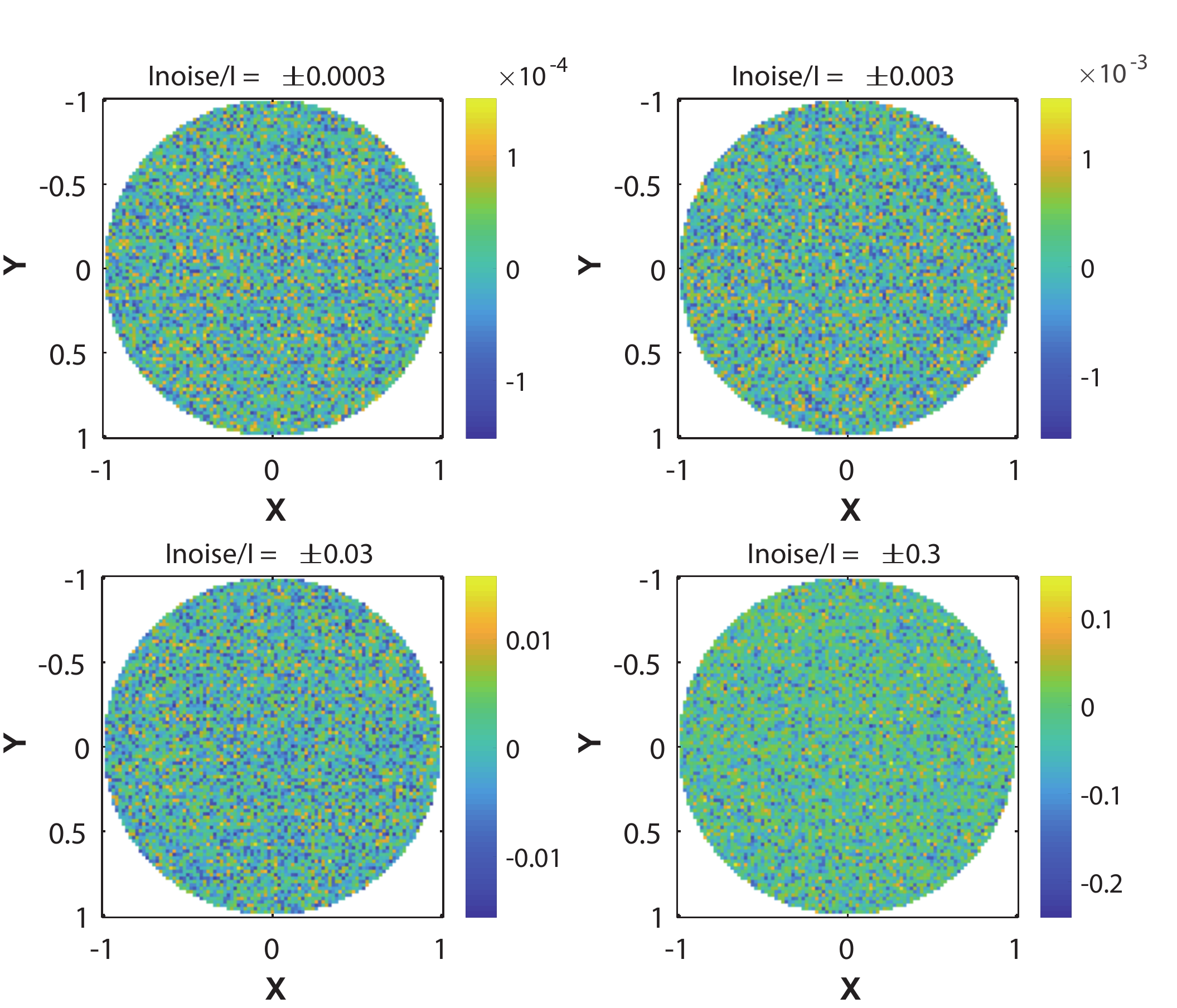}
\end{center}
  \caption{The pupil of the 
error between the true and calibrated PSG matrices as a variation of the intensity noise. The colormap displays linear relation between the input normalized intensity noise and the 
error across the pupil of the element (3,1) of the 4$\times$4 PSG error matrix $\Delta W_{3,1 (\beta=9)}$ .The pupil is scaled to have a radius of 1. }
      \label{Fig:w31}
      \end{figure}

To verify the validity of our numerical tool for the tolerance analysis, we simulate the PSG error $\Delta W$ matrix as a variation of the intensity error. The 
stochastic noise of the calibration sample $\Delta (M_i^{\rm DP})$ in Eq.~\eqref{hisim} is idealized to be 0. Modeling results show that the error across the pupil of one element of the 4$\times$4 PSG matrix $\Delta W_{3,1 (\beta=9)}$ increases linearly with the intensity noise as expected as shown in Fig.~\ref{Fig:w31}. The intensity noise normalized by the intensity of the PSA and PSG varies from $\pm 0.0003,\ \pm 0.003,\ \pm 0.03$ to $\pm 0.3$ and they are modeled as statistically the same at each pixel. Settings of the PSG and PSA are listed in Tab.~\ref{Tab:WAcoef} with the laser source having the electrical field of $E_{in} = [1; 1]/\sqrt{2}$.


\section{Mathematics for the OZP}
For non-depolarizing samples, each pixel inside the pupil of a SUT can be characterized by a 2$\times$2 complex Jones matrix. The Jones matrix at each pixel of a pupil is decomposed to two scalars and three Jones matrices to have clear physical interpretation \cite{Geh2007}, as
\begin{equation}
J\approx t e^{i\Phi_{\rm global}} J_{\rm pol}(d,\gamma) J_{\rm rot}(\alpha) J_{\rm ret}(\phi,\beta).
\label{J5}
\end{equation}
The two scalars are a transmission $t$ and a global phase $\Phi_{\rm global}$, while the three Jones matrices are for a partial polarizer $J_{\rm pol}$, a rotator $J_{\rm rot}$ and a retarder $J_{\rm ret}$, in the form of 
\begin{eqnarray}
J_{\rm pol}(d,\gamma) & = &
\begin{bmatrix}
\cos\gamma & -\sin\gamma \\
\sin\gamma & \cos\gamma  
\end{bmatrix}
\begin{bmatrix}
1+\frac{d}{2} & 0 \\
0 & 1-\frac{d}{2}  
\end{bmatrix}
\begin{bmatrix}
\cos\gamma & \sin\gamma \\
-\sin\gamma & \cos\gamma  
\end{bmatrix} \nonumber\\
&=&
\begin{bmatrix}
1+\frac{d}{2}\cos 2\gamma & \frac{d}{2}\sin 2\gamma\\
\frac{d}{2}\sin 2\gamma & 1-\frac{d}{2}\cos 2 \gamma
\end{bmatrix},\label{Jpol}\\
J_{\rm rot}(\alpha) & = &
\begin{bmatrix}
\cos\alpha & -\sin\alpha \\
\sin\alpha & \cos\alpha 
\end{bmatrix},\\
J_{\rm ret}(\phi,\beta) & = &
\begin{bmatrix}
\cos\beta & -\sin\beta \\
\sin\beta & \cos\beta 
\end{bmatrix}
\begin{bmatrix}
e^{\frac{i\phi}{2}} & 0 \\
0 & e^{\frac{-i\phi}{2}} 
\end{bmatrix}\nonumber\\
& & \begin{bmatrix}
\cos\beta & \sin\beta \\
-\sin\beta & \cos\beta  
\end{bmatrix}\\
& = &
\begin{bmatrix}
\cos\frac{\phi}{2}+i\sin\frac{\phi}{2}\cos 2 \beta & i\sin\frac{\phi}{2}\sin 2\beta \nonumber\\
i\sin\frac{\phi}{2}\sin 2 \beta & \cos\frac{\phi}{2} - i\sin\frac{\phi}{2}\cos 2 \beta
\end{bmatrix}\label{Jret},
\end{eqnarray}
where the diattenuation $d$ is the transmittance amplitude difference between the two orthogonal polarization eigenstates, and the retardance $\phi$ describes the retardance difference between those. Azimuthal angles $\gamma$ and $\beta$ determine the directions of the partial polarizer and the retarder, respectively.

The Jones matrices for the partial polarizer and the retarder can be further expanded by the OZP \cite{Tilmann2008,Ruoff2009,Ruoff2010,Meng2019}, similar to the Zernike polynomials wavefront expansion. Orientation describes the polarization directions of diattenuation and retardance. Orthogonal unity orientor matrices in polar coordinate $(r,\omega)$ with radius $r=1$ are given by
\begin{align}
\begin{split}
& {\bf O}_0^m(\omega)  = 
\begin{bmatrix}
\cos m\omega & \sin m\omega\\
\sin m \omega & -\cos m \omega
\end{bmatrix}, \\
& {\bf O}_1^m(\omega) = 
\begin{bmatrix}
\sin m \omega & -\cos m \omega\\
-\cos m\omega & -\sin m\omega
\end{bmatrix}, 
\label{Ori}
\end{split}
\end{align}
where $m$ indexes azimuthal degree $\omega$.

The Jones pupil matrix for partial polarization and retardance can be divided into a diagonal matrix with equal values of non-zero elements and the rest term. The rest term of the pupil is expanded by a sum of the OZP with their coefficients $coe_j$ to the $j$\textsuperscript{th} order. In polar coordinate $(r,\omega$), each pixel on the pupil of the partial polarizer described by Eq.~\eqref{Jpol} and the retarder described by Eq.~\eqref{Jret} are expanded by the OZP as
\begin{align}
&  J_{\rm pol}(d,\gamma,r,\omega) \nonumber \\
& =  I(r,\omega)+\frac{d(r,\omega)}{2}
\begin{bmatrix}
\cos 2\gamma(r,\omega) & \sin 2\gamma(r,\omega)\\
\sin 2\gamma(r,\omega) & -\cos 2 \gamma(r,\omega)
\end{bmatrix}\nonumber\\
& \approx  I(r,\omega) + \sum_j coe_j {\bf OZ}_j(r,\omega) \label{Jpoloz}
\end{align}

\begin{align}
& J_{\rm ret}(\phi,\beta,r,\omega) =
\cos\frac{\phi(r,w)}{2}I(r,\omega) \nonumber\\ 
& +  i\sin\frac{\phi(r,\omega)}{2}
\begin{bmatrix}
\cos 2\beta(r,\omega) & \sin 2\beta(r,\omega)\\
\sin 2\beta(r,\omega) & -\cos 2 \beta(r,\omega)
\end{bmatrix}\nonumber\\
& \approx  \cos \frac{\phi(r,\omega)}{2}I(r,\omega)+i\sum_j coe_j {\bf OZ}_j(r,\omega), \label{Jretoz}
\end{align}
where the approximation $\sin\frac{\phi(r,\omega)}{2}\approx\frac{\phi(r,w)}{2}$ is used. The term {\textbf{OZ}$_j(r,\omega)$ is further decoupled into a position ($r$) dependent term and an orientor matrix depending on the azimuths $\omega$, as
\begin{align}
& {\bf OZ}_j(r,\omega)  =  {\bf OZ}_{n,\epsilon}^m (r,\omega)=R_n^m(r){\bf O}_\epsilon^m(\omega) \\
& R_n^m(r)  =  \sum_{s=0}^{(n-|m|)/2}\frac{(-1)^s(n-s)!}{s!\Big(\frac{n+m}{2}-s\Big)!\Big(\frac{n-m}{2}-s\Big)!}r^{n-2s},
\end{align}
where $n$ indexes the highest power in radial direction and $\epsilon$=0, 1 for the 2 orientor matrix in Eq.~\eqref{Ori}. The order label $j$ represents combinations of the OZP indices $m,n,\epsilon$ with the relation $n-m=2l,\ l=0,1,...n,\ n\in \mathbb{Z}^+$. Corresponding relation between $j$ and $m,n,\epsilon$ up to the first 16 terms of the OZP is displayed in Tab.~\ref{Tab:OZPj}.

\begin{table}[h!]
\caption{The relation between $j$ and $m,n,\epsilon$ up to the first 16 terms of the OZP}
      \begin{tabular}{ccccccccccccccccc}
        \hline
         j & 1  & 2 & 3 & 4 & 5 & 6 & 7 & 8 & 9 & 10 & 11 &12 & 13 & 14 & 15 & 16 \\ \hline
         m & 0 & 0 & 1 & -1 & 1 & -1 & 0 & 0 & 2 & -2 & 2 &-2 & 3 & -3 & 3 & -3 \\
 n & 0 & 0 & 1 & 1 & 1 & 1 & 2 & 2 & 2 & 2 & 2 & 2 & 3 & 3 & 3 & 3 \\
  $\epsilon$ & 0 & 1 & 0 & 1 & 0 & 1 & 0 & 1 & 0 & 1 & 0 & 1 & 0 & 1 & 0 & 1 \\\hline        
      \end{tabular}
      \label{Tab:OZPj}
\end{table}

The OZP expansion of diattenuation and retardance are approximations. To test the accuracy of these approximations, we use the first 72 orders of the OZP. A to-be OZP expanded and reconstructed diattenuation pupil consists of 4 pupils of elements, among which two pupils are independent. We label the upper-left element in the matrix as $J{\rm input}^{\rm dia}_{\rm xx}=\frac{d}{2}\cos 2 \gamma$ and that in the upper-right as $J{\rm input}^{\rm dia}_{\rm xy}=\frac{d}{2}\sin 2 \gamma$. Similarly, two independent matrix elements for retardance are  $J{\rm input}^{\rm ret}_{\rm xx}=\sin\frac{\phi}{2}\cos 2 \beta$ and  $J{\rm input}^{\rm ret}_{\rm xy}=\sin\frac{\phi}{2}\sin 2 \beta$. As shown in Fig.~\ref{Fig:app_ozp}, the pupils of the two independent elements in diattenuation or retardance matrices are compared between the reconstruction from the coefficients to the OZP and the inputs. Two independent elements of the matrix $\sum_j coe_j {\bf OZ}_j(r,\omega)$ are reconstructed from the OZP coefficients, as
\begin{eqnarray}
J{\rm reconst}_{\rm xx}(r,\omega) = \sum_{j=1}^{72} coe_j R_n^m(r)\cos(m\omega)\label{Jrecxx}\\
J{\rm reconst}_{\rm xy}(r,\omega) = \sum_{j=1}^{72} coe_j R_n^m(r)\sin(m\omega).
\label{Jrecxy}
\end{eqnarray}

\begin{figure}[h!]
\begin{center}
\includegraphics[width=0.6 \columnwidth]{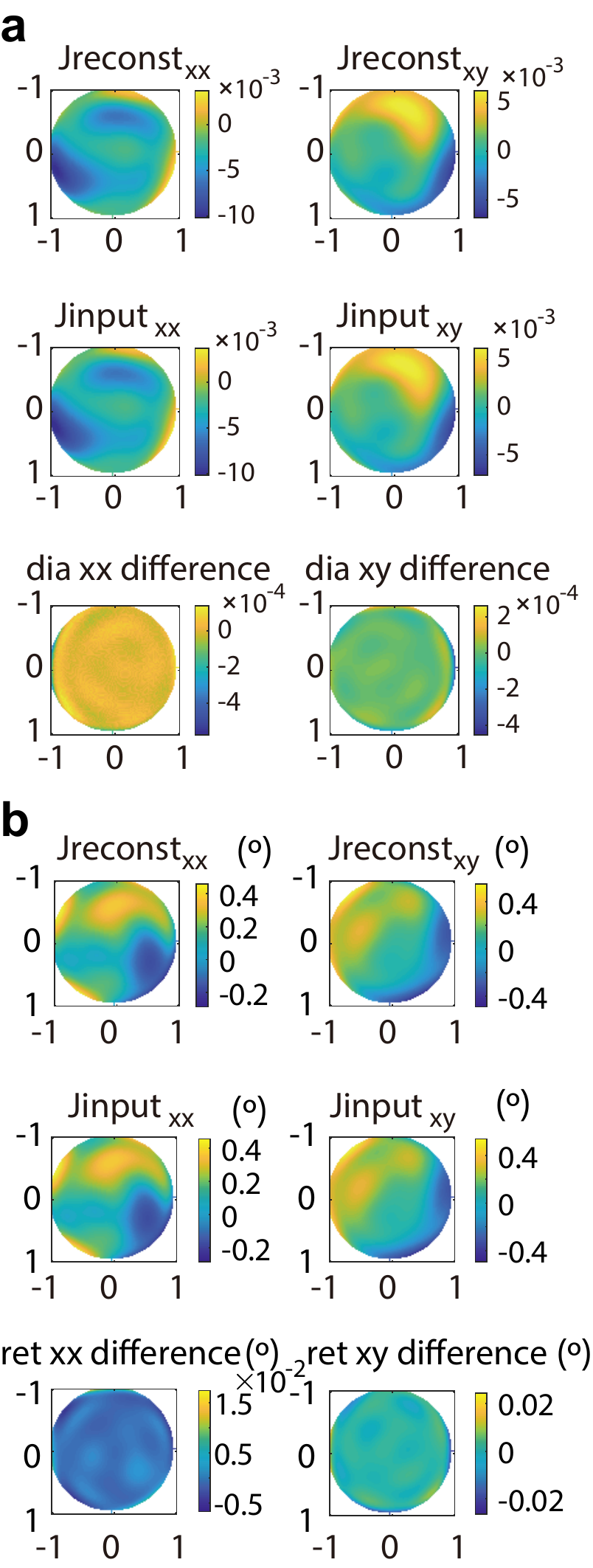}
\end{center}
  \caption{Accuracy of the Jones pupil reconstructed from the first 72 terms of the OZP.  {\bf a} Two independent elements of the Jones matrix are made for comparison, which are the upper-left element in the matrix labeled with subscript 'xx' and the element in the upper-right labeled as 'xy'. The input pupil for diattenuation is from Fig.~\ref{Fig:decomp_meas}a. {\bf b}  The input pupil for retardance is from  Fig.~\ref{Fig:decomp_meas}d. The difference between the reconstructed and the input original pupil based on the first 72 orders of the OZP is around 1/10 of either of them.}
      \label{Fig:app_ozp}
      \end{figure}

Differences from the input $J{\rm input}$ and reconstructed $J\rm reconst$ Jones matrix pupils are an order of magnitude less than either the input $J\rm input$ or the reconstructed $J\rm reconst$. Diattenuation and retardance pupils used for comparison comes from those in Fig.~\ref{Fig:decomp_meas}a and Fig.~\ref{Fig:decomp_meas}d, respectively. This difference is around 1/3 of that between the mean of the measured pupil and the repeatability pupil for either diattenuation (in Figs.~\ref{Fig:decomp_meas}b-\ref{Fig:decomp_meas}c) or retardance (in Figs.~\ref{Fig:decomp_meas}e-\ref{Fig:decomp_meas}f), demonstrating that the OZP expansion well represents diattenuation and retardance pupils. Therefore, errors contributed from the  reconstruction with the first 72 orders of the OZP is negligible in calculating the sensitivities for diattenuation and retardance. The reconstruction of the diattenuation and retardance pupils as well as the direction of the partial polarizer and the retarder from the two independent elements $J{\rm reconst}_{\rm xx}$ and $J{\rm reconst}_{\rm xy}$ is given by
\begin{eqnarray}
	d_{\rm reconst} & = & 2 \sqrt{J{\rm reconst}_{\rm xx}^2 + J{\rm reconst}_{\rm xy}^2}\\
	\phi_{\rm reconst} & = &  2 \arcsin \sqrt{J{\rm reconst}_{\rm xx}^2 + J{\rm reconst}_{\rm xy}^2}\\
	\gamma_{\rm reconst} & = & \frac{1}{2}\arctan \frac{J{\rm reconst}_{\rm xy}}{J{\rm reconst}_{\rm xx}} \nonumber\\
	& & {\rm for\ diattenuation}\label{gamma}\\
	\beta_{\rm reconst} & = & \frac{1}{2}\arctan \frac{J{\rm reconst}_{\rm xy}}{J{\rm reconst}_{\rm xx}} \nonumber \\
	& & {\rm for\ retardance}. \label{beta}
\end{eqnarray}

\bibliography{ecm_tol}

\end{document}